\documentclass[aps,prb,twocolumn,superscriptaddress, show pacs]{revtex4-1}
\usepackage{bm, amsmath,amssymb}
\usepackage{graphicx} 

\begin{document}

\title{Extended elliptic skyrmion gratings in epitaxial MnSi thin films}

\author{M. N. Wilson}
\author{E. A. Karhu}
\author{A. S. Quigley}
\affiliation{Department of Physics and Atmospheric Science, Dalhousie University, Halifax, Nova Scotia, Canada B3H 3J5}

\author{U.~K.~R\"{o}{\ss}ler}
\author{A. B. Butenko}
\author{A.~N.~Bogdanov}
\affiliation{Institut f\"{u}r Festk\"{o}rper- und Werkstoffforschung Dresden, Postfach 270016, D-01171 Dresden, Germany}

\author{M. D. Robertson}
\affiliation{Department of Physics, Acadia University, Wolfville, Nova Scotia, Canada B4P 2R6}

\author{T.~L.~Monchesky}
\affiliation{Department of Physics and Atmospheric Science, Dalhousie University, Halifax, Nova Scotia, Canada B3H 3J5}

\email[]{theodore.monchesky@dal.ca}
\thanks{}

\date{\today}

\begin{abstract}
A detailed investigation of the magnetization processes in epitaxial MnSi thin film reveals the existence of elliptically distorted skyrmion strings that lie in the plane of the film. We provide proof that the uniaxial anisotropy stabilizes this state over extended regions of the magnetic phase diagram.  Theoretical analysis of an observed cascade of first-order phase transitions is based on rigorous numerical calculations of competing chiral modulations, which shows the existence of helicoids, elliptic skyrmions, and cone phases.
\end{abstract}


\maketitle

\section{introduction}
The recent synthesis of epitaxial thin films of MnSi, \cite{Karhu:2010prb, Karhu:2011prb, Karhu:2012prb}
FeGe, {\cite{Huang:2012prl}  and (Fe,Co)Si \cite{Porter:2012prb} with a cubic B20 crystal structure represents an innovation that will facilitate the control of complex magnetic textures in chiral magnetic systems in ways that were previously inaccessible to bulk materials. 
In particular, a set of MnSi films \cite{Karhu:2011prb, Karhu:2012prb}
with in-plane magnetization enabled the observation and manipulation of
nonlinear spatial modulations with a fixed handedness 
(\textit{helicoids}), theoretically predicted by Dzyaloshinskii
more than forty years ago. \cite{Dzyaloshinskii:1964jetp}
Two-dimensional hexagonal lattices composed of localized solitonic
cores with an axisymmetric spin arrangement
(\textit{chiral skyrmions}) \cite{Bogdanov:1989fk, Bogdanov:1994jmmm}
have been identified in broad temperature and magnetic field ranges
in epitaxial FeGe films with a  skyrmion-core magnetization that is perpendicular to the surface.
\cite{Huang:2012prl} Skyrmion states also have been observed in Fe/Ir nanolayers stabilized by surface/interface induced chiral interactions.\cite{Heinze:2011fk}

In contrast, bulk MnSi and other cubic helimagnets possess a single
harmonic one-dimensional modulation (\textit{cone} phase) that corresponds to the global minimum
of the system in nearly the entire magnetically ordered region of the field-temperature
phase diagram, below the saturation field, $H_{C2}$. Within this area,
helicoids and skyrmions can exist as metastable states. 
Non-trivial magnetic ordering appears only in a small region of the phase diagram
near the ordering temperature (e.g. see Refs.~\onlinecite{Grigoriev:2006jb,Muhlbauer:2009ol,Pappas:2009bh,Wilhelm:2011prl,Onose:2012prl} and the bibliography in Ref.~\onlinecite{Wilhelm:2012jpcm}).

Chiral skyrmions are considered as promising
objects for new types of magnetic storage technologies
and for other applications.\cite{Rossler:2011jpcs,Kiselev:2011jpd}
To explore the use of non-centrosymmetric magnets in spintronic applications,
it is important to be able to create complex modulations that are stable well
below the ordering temperature.   Helicoids and skyrmion states were imaged over
extended regions of the magnetic phase diagram in mechanically thinned crystals.
\cite{Yu:2010nat, Yu:2011nm} 
This implies a crucial role of confined geometries and surface/interface induced magnetic interactions in stabilizing these textures.\cite{Butenko:2010prb, Karhu:2012prb}
Epitaxial MnSi thin films make three important contributions to this problem. 
The substrate enables control of the strain in the film, which we show stabilizes
non-trivial chiral modulations.\cite{Butenko:2010prb, Karhu:2012prb}
Secondly, epitaxial films make these extended regions accessible to a broad range
of techniques that enable us in this article to identify their structure and to show
the origin of their stability in MnSi.  Thirdly, these films stabilize complex
spin textures on a technologically relevant substrate, which opens the possibility
of engineering chiral structures.

All previous studies of extended skyrmion states reported the observation of 2D skyrmion lattices with their axis aligned perpendicular to the surface.\cite{Yu:2010nat, Yu:2011nm,Tonomura:2012nl, Huang:2012prl} We observe a grating of elliptic skyrmion strings with the core magnetizations aligned in the plane of the film where the effective field of the hard-axis uniaxial anisotropy stabilizes this state and creates the elliptic distortion depicted in Fig.~\ref{fig:HvsK}(b) and (c). 

In this paper, we present a detailed analysis of the magnetization processes in a set of epitaxial
MnSi films as a function of the uniaxial anisotropy, $K_u$, which is tuned by choice of temperature, $T$, and sample thickness, $d$.  These results, together with the calculated magnetization curves, allow us to identify the existence of an elliptic skyrmion lattice in a broad range of magnetic fields.  

\section{methods}
We grew epitaxial MnSi films on high resistivity (3-5k$\Omega$-cm) Si(111) wafers by co-deposition of Mn and Si by MBE,  as described in Ref.~\onlinecite{Karhu:2011prb}.  
Co-deposition provides large improvements in interface quality over samples grown by solid-phase epitaxy,\cite{Magnano:2006ss, Karhu:2010prb, Karhu:2011prb}  
although transmission electron microscopy (TEM) and x-ray diffraction (XRD) measurements show that the samples investigated contain a small amount of a MnSi$_{1.7}$ impurity phase, as shown in Table \ref{table:samples}.  However, previous work detected no significant modification to the magnetic properties due to these impurities.\cite{Karhu:2011prb, Karhu:2012prb}
The high quality of the films is reflected in the residual resistivity ratio (measured between $T = 299$~K and 2~K), which is above 25 for the thicker samples and is considerably larger than the value obtained for sputtered films.\cite{Huang:2012prl}  Table~\ref{table:samples} also gives the Curie temperatures ($T_C$) of the samples, which were determined from the peak in the derivative of the electrical resistivity versus temperature, and from the remanent magnetization.  The magnetization measurements were made with a superconducting quantum interference device (SQUID) magnetometer on a range of MnSi thicknesses, $18$~nm~$ < d  <  30$~nm.%

\begin{table}[!] 
\caption{\label{table:samples}
Variation in the sample quality for films of thickness $d$, as determined by the fraction of the film occupied by MnSi$_{1.7}$ impurity phase, and from the residual resistivity ratio, RRR.  The Curie temperature is also included.}
\begin{ruledtabular}
\begin{tabular}{c c c l}
$d$ (nm) & MnSi$_{1.7}$ phase (\%) & RRR  & $T_C$ (K)\\
12.8 & 4.0 & 15.4 & 41.6 \\
18.3 & 11 & 15.0 & 41.4\\
23.6 & 5.3 & 27.1 & 43.5 \\
25.4 & - & 26.8 & 42.3  \\
26.7 & $<1$ & - & 44.0  \\
29.8  & 5.6 &  25.2 & 42.8 \\
\end{tabular}
\end{ruledtabular}
\end{table}

Polarized neutron reflectometry and TEM show that the films have both left-handed and right-handed chiralities due to the inversion domains created by the growth of non-centrosymmetric films on centrosymmetric substrates.\cite{Karhu:2011prb}   In the ground state, the propagation vector $\mathbf{Q}$ of the helix points along the film normal with a reduced wavelength $L_D = 2 \pi / Q = 13.9$ nm compared to bulk.\cite{Karhu:2011prb}

This article presents magnetometry measurements with applied fields, $H$, along the in-plane MnSi$[1\overline{1}0]$ and the out-of-plane MnSi[111] directions for a complete set of temperatures below $T_C$. 
Theoretically modulated states are described within 
the standard phenomenological theory for cubic helimagnets.
\cite{Dzyaloshinskii:1964jetp,Bak:1980so}
Following Refs.~\onlinecite{Bak:1980so,Karhu:2012prb}, we write the energy density of a MnSi film as
\begin{eqnarray}
\label{Eq:w}
w (\mathbf{M}) =
\frac{c}{2} M_s^2( \nabla \mathbf{m} )^2 +
b M_s^2 \mathbf{m} \cdot 
(\nabla \times \mathbf{m}) \\ \nonumber
+ K_u (\mathbf{m} \cdot \hat{\mathbf{n}})^2
- \mu_0 \mathbf{H} \cdot \mathbf{M} 
- \frac{1}{2} \mu_0 \mathbf{H}_d \cdot \mathbf{M}\,,
\end{eqnarray}
where $\mathbf{m} = \mathbf{M}/M_s$ is a unit vector 
along the direction of the magnetization $\mathbf{M}$ ($M_s = |\mathbf{M}|$). The constants
$c=A S/(M_s^2 a^3)$ and  $b =D S/(M_s^2 a^3)$ are correspondingly the reduced values
of the spin wave stiffness, $A$, and the Dzyaloshinskii-Moriya
constant, $D$.
The spin per unit cell ($S$ = 0.8) is in units 
of $\hbar$ and $a$ = 0.4558 nm is the lattice constant.

In magnetic nanolayers, the induced uniaxial anisotropy arises
as a result of symmetry breaking
at the layers boundaries and is due to elastic strain 
imposed by a lattice mismatch between the
magnetic layer and the nonmagnetic substrate. 
\cite{deJonge:1994fk}
An inhomogeneous
distribution of the induced anisotropy across 
the film thickness may influence its magnetic
properties and even stabilize
\textit{twisted} states, as shown theoretically in Refs.~\onlinecite{Thiaville:1992jmmm,Bogdanov:2003prb}.
However, these effects 
are expected  only for certain
relations between material and external parameters. 
\cite{Thiaville:1992jmmm,Bogdanov:2003prb}
This allows us to combine all possible sources 
of induced uniaxial anisotropy into
a single effective ``volume" contribution with constant $K_u$
in Eq.~(\ref{Eq:w}) (e.g. Ref.~\onlinecite{deJonge:1994fk} 
and the discussion in Ref.~\onlinecite{Karhu:2012prb}).
Model (\ref{Eq:w}) also includes the Zeeman energy 
from the applied magnetic field, $\mathbf{H}$, and
the demagnetization energy with stray field 
$\mathbf{H}_d$. \cite{Hubert:1998}

Minimization of energy functional (\ref{Eq:w}) enables a calculation of the magnetization as a function of in-plane field in all the relevant magnetic states and yields the \textit{equilibrium} state of the system, as well as regions of \textit{metastability}.  This approach has been effective in describing non-equilibrium phenomena in bulk magnets (see Ref.~\onlinecite{Barharkhtar:1988uf} and the references therein), and allows us to explain the observed non-equilibrium hysteretic behavior. 
The magnetic phase diagram and $M(H)$ curves were calculated according to Eq.~(1) with periodic boundary conditions, assuming a constant $M_s$, where surface effects were only included in the effective volume contribution, $K_u$.  Continuum states were determined on grids with variable spacings using a finite difference method and relaxation using simulated annealing as described in Ref.~\onlinecite{Rossler:2006nat}. 

The uniaxial anisotropy is determined from the in-plane and out-of plane saturation fields  ($H_{sat}^{\|}$ and $H_{C2}^{\perp}$ respectively). The saturation field is determined from a minimum in the $d^2M/dH^2$ curve obtained from the SQUID data.  In cases where there is hysteresis in the value of the saturation field, we average the two values for $H_{sat}^{\|}$ measured on increasing and decreasing field sweeps.  These critical fields, together with the saturation magnetization, $M_s$, enable a determination of the uniaxial anisotropy by using the methods outlined in Ref.~\onlinecite{Karhu:2012prb}, which have been successfully applied to MnSi  and FeGe thin films.\cite{Karhu:2012prb, Huang:2012prl}  The analysis  also gives us the effective stiffness due to the Dzyaloshinskii-Moriya interaction, $K_0 =  A Q^2 M_s / (2 g \mu_B) $, and the effective field $H_D = 2 K_0 / M_s$.  
In this analysis, we use the low temperature measurement of $L_D = 13.9$ nm, \cite{Karhu:2011prb} and argue that this is a reasonable estimate since $L_D$ depends only weakly on field and temperature.\cite{Fak:2005aa} Furthermore, the values for $K_u$ and $K_0$ extracted from the analysis vary only weakly with the choice of Q, as can be seen from Eq.~(4) of Ref.~\onlinecite{Karhu:2012prb}.

\section{results}

\begin{figure*}[]
\centering
\includegraphics[ width=17.5 cm]{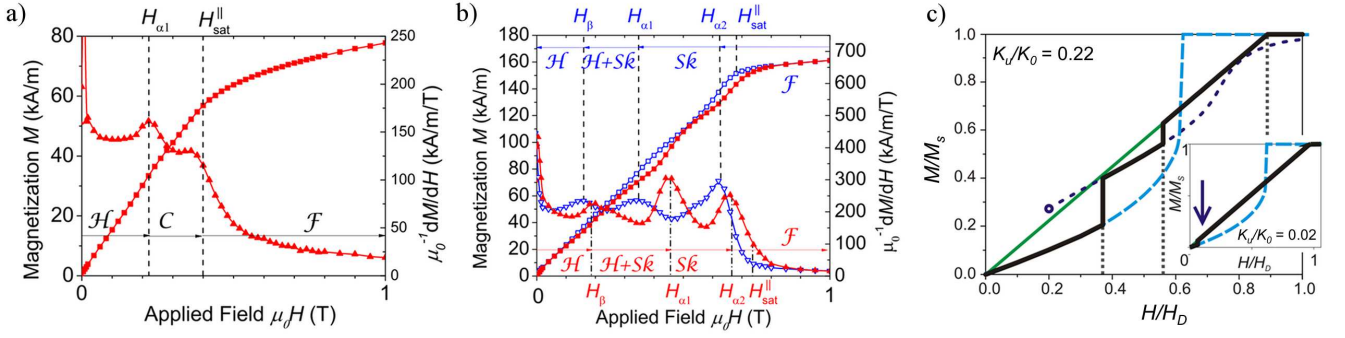}
\caption{ (color on-line) (a) and (b) show $M-H$ curves (squares) and $dM/dH$ (triangles) for a $d = 26.7$~nm film for increasing (red filled-points) and decreasing (blue open-points) magnetic fields.  (a) $T = 42$~K and $K_u/K_0 = 0.02$, and (b)  $T = 15$~K and $K_u/K_0 = 0.22$. The elliptic skyrmion $(\mathcal{S}k)$, helicoid $(\mathcal{H})$ elliptic cone $(\mathcal{C})$, and ferromagnetic $(\mathcal{F})$ regions are labeled. (c) Numerical calculations of $M(H)$ for $K_u/K_0 = 0.22$. The dashed light-blue line represents the helicoid phase, solid green line represents the conical phase, dotted navy line represents the skyrmion phase, and the thick black line follows the most energetically favorable state.  The inset shows calculations for $K_u/K_0 = 0.02$. }
\label{fig:MvsH}
\end{figure*}

Magnetization measurements at the higher temperatures presented in this paper reveal additional peaks that are not present at low temperature in MnSi thin films.  The appearance of these features and their temperature dependencies reveal the nature of different magnetic transitions in these films.   Below $H_{sat}^{\|}$, we find that the field $H_{\alpha}$ reported in Ref.~\onlinecite{Karhu:2012prb} is in fact two transitions that we label, $H_{\alpha 2}$ and $H_{\alpha 1}$, which are present in addition to the weaker transition observed at lower fields, $H_{\beta}$.

\begin{figure}[]
\centering
\includegraphics[width= 8 cm]{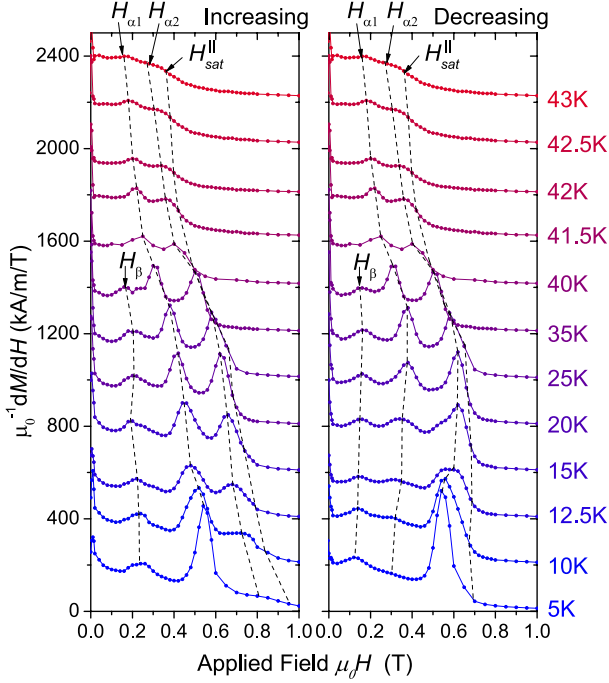}
\caption{(color on-line) Static susceptibility in increasing and decreasing field sweeps for the 26.7-nm MnSi film at various temperatures. Consecutive curves are offset by 200 kA/m/T for clarity. Dashed lines show the transition fields selected at each temperature. }
\label{fig:EK289waterfall}
\end{figure}

The results of this paper are derived by comparing the hysteresis loops and measurements of the static magnetization,  $dM/dH$,  (shown in Figs.~\ref{fig:MvsH}(a),~\ref{fig:MvsH}(b) and \ref{fig:EK289waterfall}) to the numerical calculations in Fig.~\ref{fig:MvsH}(c).
Hysteretic peaks in $dM/dH$ are clear signatures of first-order processes that arise in a number of different magnetic systems.\cite{Barharkhtar:1988uf}  In the case of chiral magnets, first-order transitions originate from the energy barriers created by the difference in topology between competing magnetic states,\cite{Wilhelm:2012jpcm} and occur via a nucleation process, as imaged in FeGe and (Fe,Co)Si.\cite{Yu:2011nm,Yu:2010nat} Measurements of these transitions over a range of $H$, $d$ and $T$ produce a thorough interrogation of the field-anisotropy phase diagram in Fig.~\ref{fig:HvsK}. 
Taken together, the results demonstrate a crossover in behavior just below $T_C$ that coincides with the stabilization of elliptic skyrmions at higher anisotropies, as shown in Fig.~\ref{fig:HvsK}. 

The calculated magnetization curves in Fig.~\ref{fig:MvsH}(c)
indicate qualitatively different magnetization processes for anisotropies below and above
the triple point ($ K_{\mathrm{tp}} =0.12 K_0$).
The inset of Fig.~\ref{fig:MvsH}(c) shows that as the field is increased for samples with $K_u  <  K_{\mathrm{tp}}$, the system evolves via a first-order transition from a helicoid into an elliptic cone phase (indicated by the arrow), and then via a second-order transition from the cone phase into the saturated state.
For $ K >K_{\mathrm{tp}}$, the skyrmion phase 
exists in an interval of magnetic
fields between the helicoid and cone phases and is separated by first-order transitions
(Fig.~\ref{fig:MvsH}(c)). 
For helicoids with a propagation vector along a hard uniaxial axis, the magnetization rotates in the easy-plane, and therefore is not influenced by the uniaxial anisotropy.
As a result, the equilibrium solutions 
for the helicoids calculated in this paper coincide with those derived by Dzyaloshinskii for an isotropic helical magnet in a transverse field.\cite{Dzyaloshinskii:1964jetp}
The magnetic field stretches the wavelength of the helicoid and causes it to diverge 
at the critical field $H_{h} =(\pi^2/16)H_D = 0.62 H_D$ where it transforms into the saturated phase.
\cite{Dzyaloshinskii:1964jetp}
Near the critical field, the helicoid dissolves into a series of isolated $360^\circ$ domain walls,\cite{Dzyaloshinskii:1964jetp} as represented by the upward turn in the calculated helicoid
$M(H)$ curve (dashed light-blue line in Fig.~\ref{fig:MvsH}(c)).
Figure~\ref{fig:HvsK}(c) shows the calculated equilibrium distributions
of the magnetization component along the skyrmion axis 
for $K_u/K_0$ = 0.22 and different values of the applied field.
Elliptic skyrmion lattices consist of repulsive localized cores, similar to  axisymmetric solutions in perpendicular magnetized helimagnets.\cite{Bogdanov:1994jmmm} Their localization 
gradually increases with increasing field. Finally, the lattice transforms into a set of isolated
skyrmions at the critical field. 

Three different behaviors are observed in the MnSi thin films in three temperature ranges.
Figure~\ref{fig:MvsH}(a) is representative of regions of low $K_u$,  where there is only one prominent peak in $dM/dH$, located at $H_{\alpha 1}$.   For the 26.7-nm thick MnSi layer, this region exists between $T = 42$~K and $T_C = 44$~K.  The numerical calculation of $M(H)$ in the inset of  Fig.~\ref{fig:MvsH}(c) is very similar to this data.  The calculated equilibrium $M(H)$ (shown in black) shows a first-order transition at a low field (indicated by the arrow) from a helicoid to a elliptically distorted conical phase, which we attribute to the peak in $dM/dH$ in Fig.~\ref{fig:MvsH}(a).  The calculated conical phase has a constant susceptibility, which explains the high-susceptibility plateau above $H_{\alpha 1}$ in Fig.~\ref{fig:MvsH}(a).  Lastly, the calculation in the inset shows a second-order transition from the conical phase into the ferromagnetic state near $H = H_D$.  This is consistent with Fig.~\ref{fig:MvsH}(a) and is further confirmation that an elliptic cone phase exists between $H_{\alpha 1}$ and $H_{sat}^{\|}$ for this temperature range.

The qualitative differences between Figs. \ref{fig:MvsH}(a) and (b) indicate a crossover in behavior.   When we increase the anisotropy by decreasing the temperature, we observe that a peak at $H_{\alpha 2}$ replaces the plateau in $dM/dH$ (as seen in Fig.~\ref{fig:EK289waterfall} ), which signals the appearance of another phase. 
 At intermediate temperatures between $T = 15$~K and 40~K, 
the amplitude of the peaks in $dM/dH$ at
$H_{\alpha 1}$ and $H_{\alpha 2}$ are approximately equal in strength. The peak at $H_{\alpha 2}$ indicates that the cone phase has nearly vanished, and the system evolves via a first-order magnetic phase transition into a ferromagnetic state.   On the increasing field branch, $H_{\alpha 1} = 0.56 H_D = 0.46$~T,  and $H_{\alpha 2} = 0.86 H_D = 0.67$~T.    A comparison between the calculated $M(H)$ in Fig.~\ref{fig:MvsH}(c) with the data in Fig.~\ref{fig:MvsH}(b), measured in an increasing field from $H_{\beta}$ to $H_{\alpha 1}$, indicates that the system is in a metastable helicoid state 
 (dashed light-blue line) 
since this phase disappears close to $H_h/H_D = 0.62$ where helicoids are no longer stable objects. This claim is strongly supported by the $H-K_u$ phase diagram (Fig.~\ref{fig:HvsK}), as discussed below.
The first-order transitions in and out of the phase bounded by $H_{\alpha 1}$ and $H_{\alpha 2}$ indicate a difference in topology between this state and the neighboring ferromagnetic and helicoid states, which leaves the elliptic skyrmion phase as the only possible explanation.  The shape of the metastable skyrmion $M(H)$ curve shown by the dotted navy line Fig.~\ref{fig:MvsH}(c) agrees nicely with the experimental data in this field range, and the appearance of hysteresis at $H_{\alpha 1}$ and $H_{\alpha 2}$ is further evidence that metastable states are formed.

As the temperature drops below 15~K, the analysis is complicated by the kinetics that play an increasingly important role in the magnetization process.  Figure \ref{fig:EK289waterfall} shows that $H_{\alpha 1}$ dominates the susceptibility for increasing field, whereas the $H_{\alpha 2}$ peak is much stronger than $H_{\alpha 1}$ for decreasing field sweeps.  The high-susceptibility plateau above $H_{\alpha 1}$ in the 10 K data in Fig.  \ref{fig:EK289waterfall} provides evidence that the conical phase reappears in the increasing field branch of the hysteresis loop.  On the decreasing branch, the evolution of the magnetic phase is similar to the intermediate temperature range with a transition from a ferromagnet to the elliptic skyrmion grating at $H_{\alpha 2}$.

\begin{figure}[]
\centering
\includegraphics[width= 8 cm]{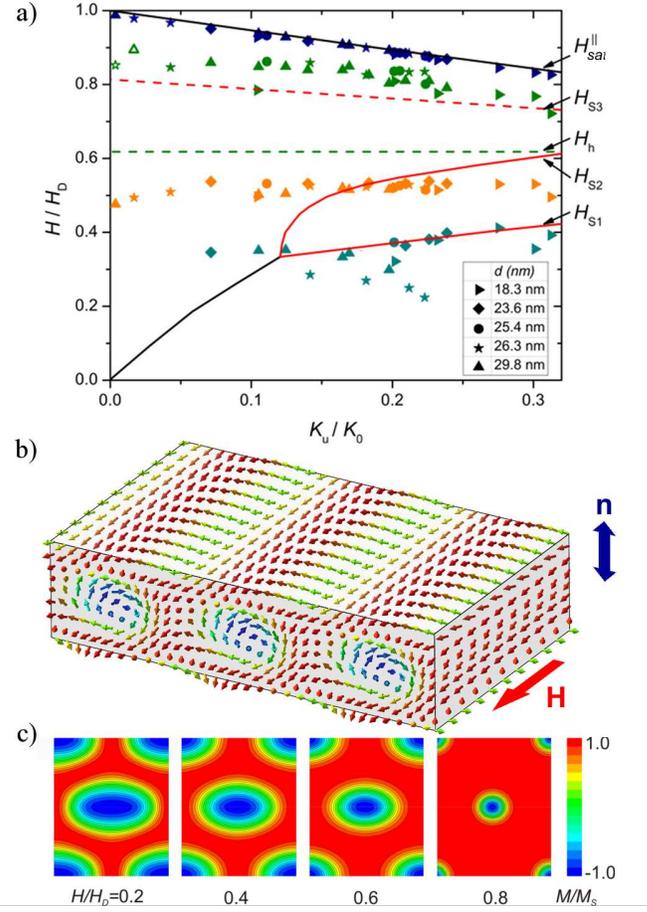}
\caption{ (Color online) (a) The critical field data for five different film thicknesses collapses into four groups that correspond in increasing field to $H_{\beta}$ (turquoise), $H_{\alpha 1}$ (orange), $H_{\alpha 2}$ (green) and $H_{sat}^{\|}$ (navy).   The two open symbols correspond to temperatures where plateaus in $dM/dH$ exist in place of peaks. $H_{S1}$ and $H_{S2}$ bound the region of stable elliptic skyrmions. The dashed-lines define the boundary for metastable helicoids, $H_h$  and elliptic skyrmions, $H_{S3}$. 
(b) Elliptic skyrmion lattice produced by an in-plane magnetic field, $\mathbf{H}$, and a uniaxial hard-axis along $\mathbf{n}$. 
(c) The equilibrium distribution of the magnetization component along the skyrmion axis calculated for  $K_u/K_0$ = 0.22 and different values of the applied field.  For  $H/H_D > 0.8$ the lattice transforms into a set of isolated skyrmions. }
\label{fig:HvsK}
\end{figure}

The construction of the anisotropy-field phase diagram paints a clear picture of the evolution of the magnetic texture in an applied field.  
To compare with the theoretical phase diagram, we avoid the more complicated behavior below $T = 15$ K.  The values for the critical fields shown in Fig.~\ref{fig:HvsK}(a) are obtained from the average of the critical fields extracted from increasing and decreasing field scans.   However, due to the small hysteresis for $T \geq 15$ K, there is little difference between the two branches of the $M-H$ loops.
Figure \ref{fig:HvsK}(a) shows that the values for $H_{\alpha 1}/H_D$ collected over a range of temperatures and sample thicknesses collapse onto a single line that is nearly independent of anisotropy and is close in value to the reduced field $H_h / H_D= 0.62$ where the calculated magnetic susceptibility of the helicoid phase diverges (see the dashed light-blue line in Fig.~\ref{fig:MvsH}(c)). This is strong evidence that $H_{\alpha 1}$ is a transition from a helicoid state.
The values of $H_{\alpha 2}/H_D$ also collapse nicely onto a line 
that is close to the line $H_{S3}$ that marks the field where the metastable elliptic skyrmion state disappears, the skyrmion analog to $H_{h}$.
The fact that the $H_{\alpha 2}/H_D$ data lie along a line with a slope that is close to that of $H_{S3}$ suggests that  a metastable elliptic skyrmion phase exists between $H_{\alpha 1}$ and $H_{\alpha 2}$. 
Above $H_{S3}$, the elliptic skyrmion lattice evaporates into an elliptic skyrmion gas of isolated vortices, similar to isolated vortices that have been directly observed in (Fe,Co)Si.\cite{Yu:2010nat}

\begin{figure}[]
\centering
\includegraphics[width=8 cm]{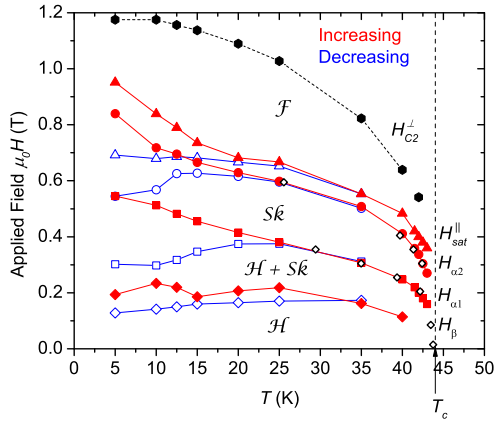}
\caption{(color online) The in-plane field $H-T$ phase diagram for a $d = 26.7$~nm MnSi film for increasing (filled red) and decreasing field (open blue) 
extracted from Fig.~\ref{fig:EK289waterfall}.  The open black diamonds are obtained from the peaks in Fig.~\ref{fig:ChivsT}.  The out-of-plane saturation fields, $H_{C2}^{\perp}$, that we used to calculate the anisotropy are also included (black hexagons).  The phases at intermediate temperatures are labeled with $\mathcal{H}$ (helicoid), $\mathcal{S}k$ (elliptic skyrmion), and $\mathcal{F}$ (ferromagnet).
 }
\label{fig:EK289phase}
\end{figure}

The region between $H_{S1}$ and $H_{S2}$ in Fig.~\ref{fig:HvsK}(a) is the region where the elliptic skyrmions are thermodynamically stable.  $H_{\beta}$ coincides approximately with this lower bound.  However, it shows much more scatter from sample to sample, as compared to $H_{\alpha 2}$ and $H_{\alpha 1}$.   This is explained by the nucleation of isolated elliptic skyrmions at defect sites, but the above analysis shows that metastable helicoids persist and create a mixed elliptic skyrmion / helicoid state, similar to the skyrmion clusters imaged inside the helicoidal phases of (Fe,Co)Si and FeGe.\cite{Yu:2010nat,Yu:2011nm}  
A comparison between the scatter in $H_{\beta}$ and the relative fraction of impurity phase in Table \ref{table:samples} shows a lack of correlation, which suggests that the MnSi$_{1.7}$ inclusions are not responsible for the nucleation of skyrmions. 
Another defect that is present in all MnSi thin film samples is the grain boundaries between the left-handed and right-handed crystals.  In FeGe polycrystals, half-skyrmion like features are observed at the boundary between crystals of opposite chirality.\cite{Yu:2011nm} Given the small $dM/dH$ at $H_{\beta}$, the elliptic skyrmions would occupy a relatively small fraction of the total sample.  If skyrmions with a diameter $L_D$ are nucleated at the grain boundaries, where the grains are of the order of 500 nm in diameter in our films,\cite{Karhu:2011prb} then the skyrmion phase would occupy approximately 10\% of the film and explain the small but non-negligible size of the $H_{\beta}$ feature.

The critical fields extracted from Fig.~\ref{fig:EK289waterfall} are also collected in Fig.~\ref{fig:EK289phase} to present the more common $H-T$ phase diagram. This shows a strong hysteresis for all four critical fields, $H_{\beta}$, $H_{\alpha1}$, $H_{\alpha2}$, and $H_{sat}$ at low temperatures, which emphasizes the first-order nature of these processes. This diagram is similar in appearance to that of (Fe,Co)Si,\cite{Yu:2010nat} more so than that of MnSi.\cite{Tonomura:2012nl}   Since transition metal site disorder exists in (Fe,Co)Si, and the stoichiometry in our MnSi films is not perfect, this raises the question of whether atomic scale defects affect the phase diagram.  It is also interesting to compare the diagram to that of doped MnSi films, \cite{Potapova:2012prb} where Ge impurities create a lattice expansion that increases the size of the A-phase.  This result reflects the role of anisotropy in stabilizing the A-phase skyrmions and emphasizes the importance of the uniaxial anisotropy in stabilizing elliptic skyrmions over a large temperature range in MnSi thin films. 
 
As a further confirmation of the critical fields, we extracted the static susceptibility from field cooled magnetization measurements, similar to measurements of the A-phase in bulk MnSi.\cite{Bauer:2012prb} We performed a series of magnetization measurements in a fixed field while warming the samples from $T = 5$~K, and  calculated $dM / dH$ from pairs of datasets, $M(T, H_1)$ and $M(T, H_2)$, in fields that differ by $H_2 - H_1 = 0.01$~T.  The results are plotted in Fig.~\ref{fig:ChivsT}, where the field values displayed on the right are $(H_1 + H_2)/2$.  The  $dM / dH$ measured in a field $H=0.355$~T is interesting in particular, as this makes a clear cut across the helicoid and skyrmion phases (see Fig.~\ref{fig:EK289phase}).  We observe the expected peaks in $dM / dH$, and there is a clear drop in the susceptibility in the skyrmion phase that lies between $T = 25$~K and $T= 40$~K, similar to what is observed in the A-phase in bulk MnSi.~\cite{Bauer:2012prb}  The critical fields determined from peaks in $dM / dH$ vs $T$ are shown in Fig.~\ref{fig:EK289phase}, and are in excellent agreement with the values obtained from $dM / dH$ vs $H$.

\begin{figure}[]
\centering
\includegraphics[width=8 cm]{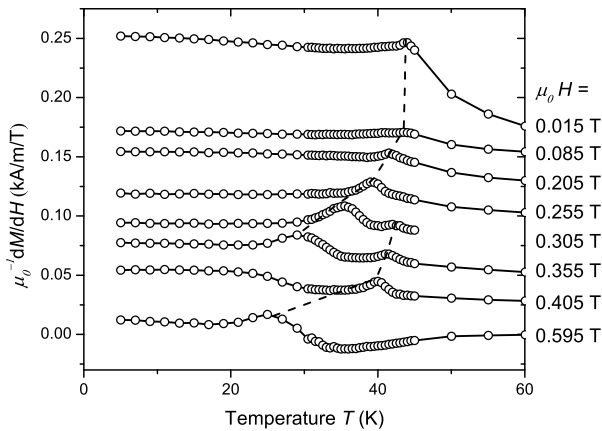}
\caption{
The temperature dependence of the static susceptibility for a 26.7 nm thick sample measured in fixed fields ranging from 0.010 T to 0.6 T.   The dashed lines show the transition temperatures selected at each field, which are plotted in Fig.~\ref{fig:EK289phase}.
}
\label{fig:ChivsT}
\end{figure}

In this paper, we use the inflection point in the magnetization curves to find estimates of $H_{sat}^{\|}$ that are required to calculate the uniaxial anisotropy.   $H_{\alpha2}$, on the other hand, corresponds to a peak in $dM / dH$ that signals a first order phase transition.  One question that arises is whether the transition $H_{\alpha2}$ is physically distinct from $H_{sat}^{\|}$, or whether the $H_{sat}^{\|}$ feature is merely a tail of the $H_{\alpha2}$ peak. This question amounts to answering whether the phases between the saturated state and the skyrmion phase are simply a mixture of these two phases, or whether a conical phase exists in this region.  To answer this question would require careful microscopy measurements, and it may very well depend on temperature.  The calculation shown in Fig.~\ref{fig:MvsH}(c) predicts a conical phase near $H_{sat}^{\|}$, but it is not certain whether the kinetics of the transitions permit the system to access this state.  However, it is worth noting that there is a small difference in the line of best fit through $H_{\alpha2}$ and $H_{sat}^{\|}$ in Fig.~\ref{fig:HvsK}(a) that may be suggesting that these two transitions have different physical origins.

\section{conclusion}

In conclusion, this paper demonstrates the presence of a unique elliptic skyrmion phase with an in-plane core magnetization.  The disappearance of this phase at low anisotropy and its retention over a large range of fields and temperatures confirms theoretical predictions that $K_u$ stabilizes elliptic skyrmions over extended regions of the phase diagram. This is an important result in the context of recent experiments that show that electric currents can displace skyrmions that are stabilized over extended regions of the phase diagram in mechanically thinned crystals, \cite{Yu:2012nc} or can manipulate the complex textures that exist in the A-phase. \cite{Schulz:2012np}  
The elliptic skyrmion gratings may provide geometric advantages for the kind of devices proposed for skyrmionic materials.\cite{Kiselev:2011jpd} The in-plane skyrmions will experience much smaller demagnetizing effects, and  furthermore, in a lithographically patterned wire with a transverse magnetic field, the grating has the advantage of permitting only one skyrmion string to span the width of the wire at a given point along its length. By stabilizing complex textures on a Si substrate over a wide range of temperatures, this material opens the opportunity for fundamental chiral spintronics experiments while the search continues for an interface-engineered Dzyaloshinskii-Moriya interaction \cite{Bode:2007nu, Heinze:2011fk} that may one day stabilize these states above room temperature.  

\begin{acknowledgements}

TLM and MNW acknowledge support from NSERC, and the support of the Canada Foundation for Innovation, the Atlantic Innovation Fund, and other partners which fund the Facilities for Materials Characterization, managed by the Institute for Research In Materials.

\end{acknowledgements}


%
\end{document}